
\magnification=\magstep1
\overfullrule=0pt

\def\L{\cal L}
\def\i{\item}
\def\dda{\sqcup\!\!\!\!\sqcap}
\def\Buildrel#1\over#2{\mathrel{\mathop{\kern0pt #1}\limits_{#2}}}
\quad
\rightline{HD--THEP--95--16}
\vskip1.5cm
\centerline{\bf BATALIN-FRADKIN-TYUTIN EMBEDDING}
\centerline{\bf OF A SELF-DUAL MODEL}
\centerline{\bf AND THE MAXWELL-CHERN-SIMONS THEORY}
\vskip1cm
\centerline{R. Banerjee\footnote*{On leave of absence from
S.N. Bose Natl. Ctr. for Basic Sc. DB-17, Sec. 1, Salt Lake, Calcutta 700064,
India} and Heinz J. Rothe}
\medskip
\centerline{Institut f\"ur Theoretische Physik}
\centerline{Universit\"at Heidelberg}
\centerline{Philosophenweg 16, D-69120 Heidelberg}
\baselineskip14pt
\vskip2cm
\centerline{\bf Abstract}
\bigskip
We convert the self-dual model of Townsend, Pilch, and Nieuwenhuizen
to a first-class system using the generalized canonical formalism
of Batalin, Fradkin, and Tyutin and show that gauge-invariant
fields in the embedded model can be identified with observables
in the Maxwell-Chern-Simons theory as well as with the fundamental
fields of the self-dual model. We construct the phase-space
partition function of the embedded model and demonstrate how a
basic set of gauge-variant fields can play the role of either
the vector potentials in the Maxwell-Chern-Simons theory or the
fundamental fields of the self-dual model by appropriate choices
of gauge.
\vfill\eject

\noindent{\bf 1. Introduction}
\bigskip
The self-dual (SD) model in 2+1 dimensions [1] has been the subject
of much discussion [2-4], because of its close connection with
the Maxwell-Chern-Simons (MCS) theory [5]. An obvious difference
between these two models is that, whereas the MCS-theory is manifestly
gauge-invariant, possessing \underbar{only} first class constraints,
the dual model is a purely second-class system. However, as was
shown in [2], the algebra of the fundamental fields in the SD-model
is identical to that of the dual field strengths in the MCS
theory. Furthermore the Schwinger defintion of the energy-momentum
tensor has an identical structure in both models, provided that
in the MCS theory it is expressed in terms of the dual field strengths.
This suggests that the SD-model and the MCS theory may just be different
gauge-fixed versions of a parent gauge theory. We shall show in this
paper that this is indeed the case. Since the MCS theory is
manifestly gauge-invariant, the natural starting point for
constructing the parent theory is the SD-model. This model can be
converted into a first-class system in an extended phase space
by following the general ideas of Batalin, Fradkin
and Tyutin [6]. Embedding second-class systems  into gauge theories
has proved to be useful within other contexts [7]. Proceeding in
this way, the gauge-invariant fields of the embedded theory are
shown in section 2 to be equivalent to the fundamental fields of the SD-model.
The \underbar{strong} self-duality constraint in the SD-model,
which does not involve the time derivative of the fields, is
lifted by the embedding procedure to a \underbar{weak} duality
constraint. This constraint turns out to be the generator of the
gauge transformations, and hence plays the role of the Gauss
law in electrodynamics. The remaining  self-duality relations are
just the equations of motion. Furthermore the involutive Hamiltonian
can be expressed in terms of the gauge-invariant fields, modulo
a term proportional to the generator of gauge transformations. This
signalizes the existence of an underlying Maxwell-type gauge
theory. One is then led in a natural way to the correspondence
of the SD model with the gauge-invariant sector of the
MCS-theory.

In section 3 we construct the phase-space path integral
representation for the embedded model. Here the role of this
model as a parent gauge theory becomes most transparent. In
the so-called unitary gauge [6] one recovers the partition
function of the SD-model, while in another judiciously
chosen gauge the partition function of the MCS theory is
reproduced. We find that in these gauges a set of basic fields
in the parent model alternatively play the role of
the fundamental fields in the SD-model and the vector potentials
in the MCS theory. We conclude the section by discussing
an alternative embedding procedure on the configuration space
path integral level by starting from a master Lagrangian
which possesses a gauge invariance in all fundamental fields.
This further illuminates the interplay between gauge invariance
and self-duality previously encountered in the Hamiltonian
formulation.
\vskip1cm
\noindent{\bf 2. Embedded Version of the SD-Model}
\bigskip
The Lagrangian of the self-dual model is given by
$${\L}_{SD}={1\over2}f^\mu f_\mu-{1\over2m}\epsilon^{\mu\nu\lambda}
f_\mu\partial_\nu f_\lambda,\eqno(2.1a)$$
and leads to the field equations
$$f_\mu-{1\over m}\epsilon_{\mu\nu\lambda}\partial^\nu
f^\lambda=0 \ , \eqno(2.1b)$$
which are usually referred to as the self-dual relations [1]. It
describes a purely second-class system with the constraints
$$\Omega_0\equiv{\pi_0}^{(f)}\approx0,\eqno(2.2a)$$
$$\Omega_i\equiv{\pi_i}^{(f)}+{1\over 2m}
\epsilon_{ij} f^j\approx0,\eqno(2.2b)$$
$$\Omega\equiv f_0-{1\over m}
\epsilon_{ij}\partial^i f^j\approx0,\eqno(2.2c)$$
and the canonical Hamiltonian
$$H_c =\int d^2x\left[
-{1\over2}f^\mu f_\mu+{1\over m}\epsilon_{ij}f^0\partial^if^j\right]
.\eqno(2.3)$$
The constraints (2.2b) are a manifestation of the symplectic
structure of the Chern-Simons three form appearing in
(2.1). Following [8] we shall implement them strongly by
using the modified Poisson brackets
$$\left\{f_i(x),f_j(y)\right\}=-m\epsilon_{ij}\delta(\vec x-
\vec y)\ , \quad\quad (\epsilon_{12}=1).\eqno(2.4)$$
All other brackets are conventional Poisson brackets. With
respect to these we are now left with only two second-class
constraints. We next convert the model described by the
Hamiltonian (2.3) and the constraints (2.2a) and (2.2c)
to a first-class system (with respect to the above mentioned
brackets) by following the general ideas of
ref. [6]. To this effect we enlarge the original
phase space by introducing a canonical pair $\alpha$ and $\pi_\alpha$.
Then a new set of first-class constraints
obtained from the original ones, (2.2a) and (2.2c), is given
by
$$\eqalign{
\Omega_0'&=\Omega_0+\alpha,\cr
\Omega'&=\Omega+\pi_\alpha.\cr}\eqno(2.5)$$
These constraints are in strong involution. A Hamiltonian
which is in involution with $\Omega_0'$ and $\Omega'$ is easily
constructed
$$H'=H_c+\int d^2x \left[{1\over 2}\pi^2_\alpha (x)-{1\over2}\partial^i\alpha
(x)\partial_i
\alpha (x)+\alpha (x)\partial_if^i(x)\right]\eqno(2.6)$$
satisfying the involutive algebra
$$\eqalign{
\left\{\Omega_0'(x),H'\right\}=&\Omega'(x),\cr
\left\{\Omega'(x),H'\right\}=&0.\cr}\eqno(2.7)$$
Note that $\Omega_0'$ and $\Omega'$ are the generators
of the infinitesimal gauge transformations
$$\eqalign{
&\left\{f^i(x),G[\theta]\right\}=-\partial^i\theta(x),\cr
&\left\{\alpha(x),G[\theta]\right\}=\theta(x),\cr
&\left\{f^0(x),G_0[\tilde{\theta}]\right\}=\tilde{\theta}(x),\cr
&\left\{\pi_\alpha(x),G_0[\tilde{\theta}]\right\}=-\tilde{\theta}(x),\cr}
\eqno(2.8a)$$
where
$$\eqalign{
&G[\theta]=\int d^3x\theta(x)\Omega'(x),\cr
&G_0[{\tilde\theta}]=\int d^3x{\tilde\theta}(x)\Omega_0'(x).\cr}
\eqno(2.8b)$$
{}From here we see that the following combinations of fields
$$\eqalign{
&F^i=f^i+\partial^i\alpha,\cr
&F^0=f^0+ \pi_\alpha,\cr}\eqno(2.9)$$
are gauge-invariant and satisfy the algebra
$$\eqalign{
&\{F^0(x),F^0(y)\}=0,\cr
&\{F^0(x),F^i(y)\}=\partial^i\delta(\vec x-\vec y),\cr
&\{F^i(x),F^j(y)\}=-m\epsilon^{ij}\delta(\vec x-\vec y).\cr} \ \eqno(2.10)$$
The algebra of the $F^\mu$ fields is identical to that of the
$f^\mu$-fields in the dual model, where all constraints have
been implemented strongly [2]. Hence this is true in any
gauge. In particular, in the unitary gauge [6], which
consists of taking the original second class constraints
(2.2a) and (2.2c) to be the gauge-fixing conditions, the
involutive Hamiltonian, the constraints and the algebra of
the (gauge-dependent) fields $f_\mu$ become identical to
those of the SD-model.

The equation of motion derived from the involutive
Hamiltonian (2.6) can be easily shown to have the
covariant form
$$F_\mu(x)-{1\over m}\epsilon_{\mu\nu\lambda}\partial^\nu
F^\lambda(x)=0 \ .\eqno(2.11)$$
This is the self-duality relation (2.1b) written in the
embedded version. The time component of the LHS of this
equation is just the generator of time independent gauge transformations.
We therefore see that the second class constraint (2.2c)
has been lifted in the embedded version to a generator
of gauge transformations, and therefore
plays a similar role as the Gauss operator in quantum
electrodynamics.  This is further illuminated by expressing
the involutive Hamiltonian (2.6) in terms of the gauge-invariant
fields $F^\mu$:
$$H'={1\over2}\int d^2x\left[F^2_0(x)+\vec F^2(x)\right]
-\int d^2xf_0(x)\Omega'(x).\eqno(2.12)$$
Although $f_0$ appears to play the role analogous to
the Lagrange multiplier $A_0$ in QED, such a correspondence
clearly does not hold for the spatial components $f_i$, as
is evident from the modified brackets (2.4), which show that $f_2$ and ${1\over
m}f_1$
form a canonical pair. Hence the Hamiltonian
(2.12) cannot be that of an ordinary Maxwell theory. However,
the gauge symmetry of the embedded model, and the fact that $F_\mu$ is
divergenceless (as
follows from (2.11)), suggest that this gauge
symmetry can be exposed in an alternative way by introducing gauge
potentials through
$$F_\mu(x)=\epsilon_{\mu\nu\lambda}\partial^\nu
 A^\lambda(x) \ .\eqno(2.13)$$
This is similar in spirit to the work in [9], where the Hopf
term could thereby be introduced in the non-linear sigma model.
Using (2.13) the Hamiltonian (2.12) takes the form
$$H'={1\over2}\int d^2x[\vec E^2+B^2]+{1\over m}
\int d^2xf_0(x)(\vec\nabla\cdot\vec E(x)+mB(x)) \ ,\eqno(2.14a)$$
where
$$\eqalign{
E^i&=\partial^iA^0-\partial^0A^i,\cr
B&=-\epsilon_{ij}\partial^iA^j.\cr}\eqno(2.14b)$$
The physical states are those which are annihilated by the
operator $\vec\nabla\cdot\vec E(x)+mB$, which is identical to the
Gauss operator of the Maxwell-Chern-Simons theory. Furthermore,
in the physical space, $H'$ is identical with the
Hamiltonian of the MCS theory with the correct commutation
relations involving $\vec E$ and $B$. This follows from the
brackets (2.10). We therefore see that the field (2.13) can be identified
with the dual field strength tensor in the MCS-theory. Hence by embedding the
second-class SD-model, we
have arrived in a systematic way at its equivalent formulation in the
gauge-invariant sector of the MCS theory with the correspondence
$$\eqalign{
&(f_0)_{SD}\longrightarrow(f_0+\pi_\alpha)_{emb}\longrightarrow
(F_0)_{MCS},\cr
&(f_i)_{SD}\longrightarrow(f_i+\partial_i\alpha)_{emb}\longrightarrow
(F_i)_{MCS},\cr}\eqno(2.15)$$
where the subscript ``$emb$'' stands for ``embedded model".

We conclude this section with some observations. The action
of the SD-model possesses no gauge symmetry, while the equations
of motion are the self-dual relations. In the embedded version
the theory has a gauge invariance but the fields
$f_\mu$ no longer satisfy an equation of motion, having the
self-dual form. The self-duality relation is recovered in
the gauge-invariant sector which involves a combination of the
basic fields. Correspondingly the (gauge invariant) dual field
strengths (2.13) in the MCS theory satisfy the self-dual
relations (2.11) (which are just the usual MCS equations
of motion written in terms of the dual fields), whereas this
is not the case for basic gauge-variant fields $A_\mu$. This illustrates
the interplay between gauge invariance and self-duality in these models.
\vskip1cm
\noindent{\bf 3. Path Integral Approach}
\bigskip
In the previous section we have studied the connection
between the SD- model and the MCS- theory within the Hamiltonian
framework by converting the second-class SD- model into a
first-class system in an extended phase space. The purpose ot
this section is twofold. We first construct the phase space
partition function of the embedded model considered in the previous
section and show how by appropriate choices of gauges one
recovers either the SD model or the MCS- theory. Thereafter we
shall consider a purely configuration space path integral
approach starting from a master Lagrangian, differing
from those previously studied in the literature [2-4,10].
This will complement the Hamiltonian analysis carried out
in the previous section.

Consider the involutive Hamiltonian (2.6) with the canonical
pairs given by $(f_0,\pi_0), (\alpha,\pi_\alpha)$ and
$(f_2,{1\over m}f_1)$. The phase space partition function is
therefore given by
$$Z=\int Df_0D\pi_0Df_iD\alpha D\pi_\alpha\delta
(\Omega_o')\delta(\Omega')\delta(G_i)
e^{i\int d^3x[\pi_0\dot f_0+{1\over m}f_1\dot f_2+\pi_\alpha
\dot\alpha-H']}\ ,\eqno(3.1)$$
where $\Omega_0'$ and $\Omega'$ are the first-class constraints
(2.5), $H'$ is the involutive Hamiltonian (2.6), and $G_i(i=
1,2)$ are two gauge-fixing functions which we shall take to
be linear in the fields.  Hence the Faddeev-Popov determinant
is trivial and has not been included. Taking the original
second-class constraints (2.2a) and (2.2c) as the unitary [6]
gauge-fixing conditions we obtain
$$Z=\int Df_\mu \delta(f_0-{1\over m}\epsilon_{ij}
\partial^i f^j)e^{i\int d^3x{\L}_{SD}},\eqno(3.2)$$
where ${\L}_{SD}$ is the Lagrangian (2.1a) of the SD-model.
This is the configuration-space partition function  which
one obtains from a conventional phase-space analysis.

Alternatively, consider the gauge
$$G_i=f_i-{1\over m}\epsilon_{ij}\partial^j f_0=0.\eqno(3.3)$$
Performing the (trivial) momentum integrals
one finds
$$Z=\int Df_\mu D\alpha\delta(f_i-{1\over m}\epsilon_{ij}\partial
^jf_0)e^{i\int d^3x{\L}},\eqno(3.4a)$$
where
$${\L}=-{1\over 4m^2}f^{ij}f_{ij}+{1\over 2m}\epsilon^{ij}f_i\partial
_0f_j+{1\over2}f^if_i
+{1\over2}\alpha\vec\nabla^2\alpha-{1\over m}\alpha \epsilon_{ij}
\partial^if^{0j} \ ,\eqno(3.4b)$$
with
$$f_{\mu\nu}= \partial_\mu f_\nu-\partial_\nu f_\mu.\eqno(3.4c)$$
The  $\alpha$- integral is readily performed. Making
use of $\partial_i f_i=0$ and $\partial_i f_0=-m\epsilon_{ij}
f^j$, following from the gauge condition (3.3), one finds
after some algebra
$$Z=\int DA_\mu\delta(A_i-{1\over m}\epsilon_{ij}\partial^jA_0(x))
e^{i\int d^3x{\L}_{MCS}}\ ,\eqno(3.5a)$$
where
$$A_\mu={1\over m} f_\mu,\eqno(3.5b)$$
and
$${\L}_{MCS}=-{1\over 4}F_{\mu\nu}F^{\mu\nu}+{m\over2}\epsilon_{\mu\nu
\lambda}A^\mu\partial^\nu A^\lambda\eqno(3.5c)$$
is the Lagrangian of the familiar Maxwell-Chern-Simons theory [5].
We now recognize that the gauge-fixing conditions
 (3.3) are equivalent to choosing the Coulomb gauge and $\nabla^2
A^0=m\epsilon^{ij}\partial_iA_j$, which is just Gauss' law in this gauge.
Hence $f_\mu$ now plays the role of the vector potential in the
MCS theory, while in the unitary gauge $f_\mu$ was the
fundamental field of the SD model. In a general gauge, on the
other hand, the gauge-invariant combination $f_\mu+\partial
_\mu\alpha$ (with $\dot\alpha$ defined through the Hamiltonian
equations of motion) plays the role of the self-dual field
in the SD model, or the dual field strength
$F_\mu=\epsilon_{\mu\nu\lambda}\partial^\nu A^\lambda$ in the MCS
theory.
\smallskip
This completes the phase space path integral analysis
starting from the involutive Hamiltonian of the embedded model.
We conclude this section with a pure configuration space path
integral analysis, which will further elucidate the interplay
between gauge invariance and self-duality referred to earlier.

Consider the following master Lagrangian
$${\L}=-{1\over4}F_{\mu\nu}F^{\mu\nu}+\epsilon_{\mu\nu\lambda}f^\mu
\partial^\nu A^\lambda-{1\over 2m}\epsilon_{\mu\nu\lambda}
f^\mu\partial^\nu f^\lambda,\eqno(3.6)$$
where $F_{\mu\nu}=\partial_\mu A_\nu-\partial_\nu A_\mu$. In
contrast to the master Lagrangians considered in refs. [2,10]
the action corresponding to (3.6) is independently gauge-invariant
under the transformations $A_\mu\to A_\mu+\partial_\mu\Lambda$,
and $f_\mu\to f_\mu+\partial_\mu\lambda$. The generating
functional for the gauge-invariant fields $\epsilon_{\mu\nu\lambda}
\partial^\nu A^\lambda$ and $\epsilon_{\mu\nu\lambda}
\partial^\nu f^\lambda$ is given by
$$Z[j,J]=\int DA_\mu Df_\mu\delta(\partial_\mu A^\mu)
\delta(\partial_\mu f^\mu)
e^{i\int d^3x[{\L}+(\epsilon_{\mu\nu\lambda}\partial^\nu
A^\lambda)J^\mu+{1\over m}(\epsilon_{\mu\nu\lambda}
\partial^\nu f^\lambda)j^\mu]}\ ,\eqno(3.7)$$
where a covariant gauge has been chosen for both the $A_\mu$
and $f_\mu$ fields. Performing first the integration over
$f_\mu$, implementing the gauge condition $\partial_\mu f^\mu=0$
by t'Hooft's method, one obtains
$$Z[j,J]=\int DA_\mu\delta(\partial_\mu A^\mu)
e^{i\int d^3x[{\L}_{MCS}+{1\over 2m}\epsilon_{\mu\nu\lambda}j^\mu\partial
^\nu j^\lambda+\epsilon_{\mu\nu\lambda}\partial^\nu A^\lambda(j^\mu+J^\mu)]} \
,\eqno(3.8)$$
where ${\L}_{MCS}$ is the Lagrangian of the Maxwell-Chern-Simons
theory (3.5c). For vanishing sources this is the partition
function of the MCS theory in a covariant gauge.

Alternatively, performing the $A_\mu$ integration in an
analogous way, one finds that
$$Z[j,J]=\int Df_\mu\delta(\partial_\mu f^\mu)
e^{i\int d^3x[{\L}_{SD}+{1\over2}J_\mu(g^{\mu\nu}-{\partial^\mu
\partial^\nu\over\dda})J_\nu+f_\mu J^\mu+\tilde f_\mu
j^\mu]}\ ,\eqno(3.9a)$$
where ${\cal L}_{SD}$ is given by (2.1a) and $\tilde f_\mu$ is
the dual of $f_\mu$:
$$\tilde f_\mu={1\over m}\epsilon_{\mu\nu\lambda}\partial^\nu f^\lambda \ .
\eqno(3.9b)$$
Notice that for vanishing sources (3.9) does not reduce to the
partition function of the SD-model, because of the non-trivial
measure. In fact, as we now show, the $f_\mu$-field appearing
in (3.9) does not satisfy the self-duality relation (2.1b) on the
level of Green functions. Differentiating in turn the equivalent
expressions (3.8) and (3.9) with respect to $J^\mu$ and $J^\nu$, $J^\mu$
and $j^\nu$, and $j^\mu,j^\nu$, one finds
$$\eqalign{
<F_\mu(x)F_\nu(y)>_{MCS}=&<f_\mu(x)
f_\nu(y)>-i\left(g_{\mu\nu}-{\partial_\mu\partial_\nu\over \dda}\right)
\delta(x-y)\cr
=&<f_\mu(x)\tilde f_\nu(y)>=<\tilde f_\mu(x)
\tilde f_\nu(y)>-{i\over m}\epsilon_{\mu\nu\lambda}\partial^\lambda
\delta(x-y)\cr}\eqno(3.10)$$
{}From (3.10) we see that, because of the presence of a non-local
propagating term, $f_\mu(x)$ cannot be identified with $\tilde f_\mu(x)$,
and therefore also not with the self-dual field appearing in (2.1).
Indeed, as we now demonstrate, it plays a role analogous to
that of the $f_\mu$-field in the embedded version of SD-model.
To this effect we express the $\delta$-function in (3.9a) as a Fourier
transform with Fourier variable $\beta(x)$. One then finds that (3.9a)
can be written in the form
$$\eqalign{&Z[j,J]=\cr
&\int Df_\mu D\beta
 e^{i\int \left\{{\L}_{SD}[f_\mu+\partial_\mu\beta] -{1\over
2}\partial_\mu\beta\partial^\mu\beta
+{1\over2}J_\mu(g^{\mu\nu}-{\partial^\mu
\partial^\nu\over\dda})J_\nu
+(f_\mu+\partial_\mu\beta) J^\mu
+{1\over m}[\epsilon_{\mu\nu\lambda}
\partial^\nu(f^\lambda+\partial^\lambda\beta)]j^\mu+\beta\partial_\mu J^\mu
\right\}}}\eqno(3.11a)$$
where
$${\L}_{SD}[f_\mu+\partial_\mu\beta]={1\over2}(f_\mu+\partial_\mu
\beta)(f^\mu+\partial^\mu\beta)
-{1\over 2m}\epsilon_{\mu\nu\lambda}(f^\mu+\partial^\mu\beta)
\partial^\nu(f^\lambda+\partial^\lambda\beta) \ .\eqno(3.11b)$$
Introducing the new fields
$$h_\mu(x)=f_\mu(x)+\partial_\mu\beta(x),\eqno(3.12)$$
and carrying out the integration over $\beta$ yields
$$Z[j,J]=\int Dh_\mu
e^{i\int d^3x\left\{{\L}_{SD}[h]+{1\over2}J_\mu J^\mu+h_\mu J^\mu
+{1\over m}(\epsilon_{\mu\nu\lambda}\partial^\nu h^\lambda)j^\mu\right\}}\
.\eqno(3.13)$$
By performing the appropriate differentiations with respect to the sources
$j^\mu$ and $J^\mu$ of the equivalent representations (3.8) and (3.13)
we obtain relations analogous to (3.10):
$$\eqalign{
<F_\mu(x)F_\nu(y)>_{MCS}=&<h_\mu(x)h_\nu(y)>_{SD}-ig_{\mu\nu}
\delta(x-y)\cr
=&<h_\mu(x)\tilde h_\nu(y)>_{SD}\cr
=&<\tilde h_\mu(x)\tilde h_\nu(y)>_{SD}-{i\over m}\epsilon_{\mu
\nu\lambda}\partial^\lambda\delta(x-y) \ .\cr}\ \eqno(3.14)$$
{}From here we can conclude that, apart from non-propagating contact
terms, the following identifications hold on the level of Green
functions
$$h_\mu(x)\leftrightarrow\tilde h_\mu(x)\leftrightarrow F_\mu(x)\ .$$
These are just the known correspondences [2,4] among the fundamental
fields in the SD-model, and the dual field strengths in the MCS
theory. The gauge-invariant field $h_\mu(x)$ is the
Lagrangian version of the gauge-invariant combinations $f_0+\pi_\alpha,
f_i+\partial_i\alpha$ in the embedded model discussed in section 2.
\vskip1cm
\noindent{\bf 4. Conclusion}
\bigskip
In this paper we have applied the general ideas of Batalin-Fradkin-Tyutin [6]
to obtain a deeper insight into the connection between the fields in
the self-dual model and those in the Maxwell-Chern-Simons theory. By
embedding the self-dual model in an extended phase space, the fields
in the gauge-invariant sector could be identified on one hand with
the dual field strengths in the MCS theory and on the other hand
with the fundamental fields of the SD-model. As a consequence, the
equations of motion for these fields (including the constraint)
are just the self-dual relations.

By studying the phase-space partition function corresponding to
the involutive Hamiltonian of the embedded model in two
particular gauges, we recovered the configuration space partition functions
of the SD-model as well as that of the MCS-theory. The $f_\mu$ field in the
embedded
model could
be identified with either the fundamental field in the SD-model
or the gauge potentials in the MCS theory. In this sense the involutive
Hamiltonian played the role of a parent Hamiltonian, and
the $f_\mu$ field, in the embedded model, that of an interpolating field.

Finally we have discussed an embedding procedure on the
configuration space path integral level,
differing from the St\"uckelberg [11] embedding, by starting from
a master Lagrangian which, in contrast to the ones
considered in the literature [2,10], has a gauge invariance
in all fundamental fields. Apart from revealing the common
origin of the SD-model and MCS theory, it provided an instructive
way of exhibiting the interplay between gauge invariance
and self-duality.
\bigskip
\centerline{\bf Acknowledgments}
\bigskip\noindent
One of the authors (R.B) would like to thank the Alexander- von- Humboldt
Foundation for providing the financial support which made this collaboration
possible.
\bigskip\bigskip
\centerline{\bf References}
\bigskip\bigskip\noindent
\i{[1]} P.K. Townsend, K. Pilch and P. van Nieuwenhuizen, Phys. Lett. {\bf
B136},
(1984), 38; ibid, {\bf B137}, (1984), 443 (Addendum).
\i{[2]} S. Deser and R. Jackiw, Phys. Lett. {\bf B139}, (1984), 371.
\i{[3]} E. Fradkin and F.A. Schaposnik, Phys. Lett. {\bf B338}, (1994), 253.
\i{[4]} R. Banerjee, H.J. Rothe and K.D. Rothe, ``On the Equivalence of the
Maxwell- Chern- Simons Theory and a Self- Dual Model'', Heidelberg Preprint
(HD- THEP- 95- 13).
\i{[5]} S. Deser, R. Jackiw and S. Templeton, Phys. Rev. Lett. {\bf 48},
(1982),
975; Ann. Phys. (NY) {\bf 140}, (1982), 372.
\i{[6]} I.A. Batalin and E.S. Fradkin, Nucl. Phys. {\bf B279}, (1987), 514;
I.A. Batalin and I.V. Tyutin, Int. J Mod. Phys. {\bf A6}, (1991), 3255.
\i{[7]} T. Fujiwara, Y. Igarashi and J. Kubo, Nucl. Phys. {\bf B341}, (1990),
695;
Y.W. Kim, S.K. Kim, W.T. Kim, Y.J. Park, K.Y. Kim and Y. Kim, Phys. Rev.
{\bf D46}, (1992), 4574; R. Banerjee, Phys. Rev. {\bf D48}, (1993), R5467;
R. Banerjee, H.J. Rothe and K.D. Rothe, Phys. Rev.
{\bf D49}, (1994), 5438; Nucl. Phys. {\bf B426}, (1994), 129; N. Banerjee,
R. Banerjee and S. Ghosh, Phys. Rev. {\bf D49}, (1994), 1996; Nucl. Phys.
{\bf B417}, (1994), 257.
\i{[8]} L.D. Faddeev and R. Jackiw, Phys. Rev. Lett. {\bf 60}, (1988), 1692.
\i{[9]} F. Wilczek and A. Zee, Phys. Rev. Lett. {\bf 51}, (1983), 2250.
\i{[10]} A. Karlhede, U. Lindstrom, M. Rocek and P. van Nieuwenhuizen, Phys.
Lett. {\bf B186}, (1987), 96.
\i{[11]} E.G. Stuckelberg, Helv. Phys. Acta {\bf 30}, (1957), 209.
\end

\end